\documentclass[12 pt]{article}

\usepackage{graphicx}
\usepackage{amsmath}

\newcommand{\modn}{\textrm{ mod}(n)}

\begin{document}

\title{The dual problems of coordination and anti-coordination on random bipartite graphs}

\author{Matthew I. Jones$^{1*}$, Scott D. Pauls$^1$, Feng Fu$^{1,2}$}
\date{%
$^1$ Department of Mathematics, Dartmouth College, Hanover, NH 03755, USA\\
$^2$ Department of Biomedical Data Science, Geisel School of Medicine at Dartmouth, Lebanon, NH 03756, USA\\ 
$^*$ Corresponding Author. Email: Matthew.I.Jones.GR@dartmouth.edu\\[2ex]%
    \today
}


\begin{titlepage}
\maketitle
\begin{abstract}
In some scenarios (``anti-coordination games''), individuals are better off choosing different actions than their neighbors  while in other scenarios (``coordination games''), it is beneficial for individuals to choose the same strategy as their neighbors. Despite having different incentives and resulting population dynamics, it is largely unknown which collective outcome, anti-coordination or coordination, is easier to achieve. To address this issue, we focus on the distributed graph coloring problem on bipartite graphs. We show that with only two strategies, anti-coordination games (2-colorings) and coordination games (uniform colorings) are dual problems that are equally difficult to solve. To prove this, we construct an isomorphism between the Markov chains arising from the corresponding anti-coordination and coordination games under certain specific individual stochastic decision-making rules. Our results provide novel insights into solving collective action problems on networks.
\end{abstract}
\end{titlepage}

\section{Introduction}

An $n$-coloring of a graph is a labeling of the vertices of the graph with $n$ different colors such that for each pair of vertices connected by an edge, the vertices have different labels. Finding $n$-colorings is a classic graph theoretic problem. However, in recent years, graph colorings have also been adopted into the field of collective dynamics to study networked coordination games \cite{kearns2006, mccubbins2009}.

For the purposes of this paper, collective action games fall into two broad categories: games where individuals coordinate to pick the same strategies (referred to as coordination games) \cite{jackson2014,belloc2019, mccubbins2020}, and games where individuals coordinate to pick different strategies (referred to as anti-coordination games) \cite{mccain2014,bramoulle2007,neugebauera2008, broere2017, broere2019}. Coordination games can often be resolved if the players are allowed to communicate, but asymmetries in anti-coordination games can make cooperation difficult and highly dependent on network structure \cite{bramoulle2007}. In general, these lead to vastly different population dynamics, but in this paper we will see that under certain circumstances, these two classes of games can be thought of as the dual problem of one another.

There is a rich history of playing games and modeling interactions on graphs as a way to examine the effects of our social structure \cite{kearns2006, bramoulle2007, broere2017, broere2019, mccubbins2020, choi2011, shirado2020, rand2011, szabo2007}.
In particular, many social coordination problems can be phrased as graph coloring problems, like time tabling and radio frequency assignments~\cite{sabar2012,park1996}. However, unlike in the purely graph theoretic context, these social problems come with the additional complication that individuals may not have complete knowledge of the population structure. A graph coloring problem in which each vertex has to choose its edge using only local information (the colors of its neighbors) introduces new complications to the classic problems, and stochastic behavior is often needed to successfully find an n-coloring of the graph \cite{shirado2017, jones2021}. Distributive graph coloring problems can be considered as one kind of anti-coordination game, where individuals are playing games with their neighbors and trying to choose different strategies, or colors. Solving the graph coloring is equivalent to finding the social optimum.

In this work, we consider the simple case of a connected, bipartite graph which always admits exactly two 2-colorings. For an omniscient observer that can view the entire graph and dictate colors to vertices, finding one of these 2-colorings is a trivial matter. However, things become more difficult when there is no central decision-maker, and instead each vertex represents an individual who must choose her own color with no information except the colors of her neighbors \cite{jones2021}. This new game, which uses local information instead of global information, has an interesting consequence: finding a 2-coloring of the graph, which models an anti-coordination game, is equivalent to getting all individuals in the graph to choose the same color, which is a coordination game. 

Thus, in the context of bipartite graphs, anti-coordination games and coordination games are dual problems, and a whole new class of coordination games where everyone wants to opt for the same can also be modeled as a graph coloring problem. We show this by defining two Markov chains \cite{gagniuc2017, durrett2018} on the space of colored graphs, one where individuals are playing the anti-coordination game and one where individuals are playing the coordination game, and showing that they are isomorphic.

\section{Theoretical Results}
\subsection{A Natural Bijection for Update Rules for 2-colorings and Uniform Colorings}
In this paper, the individuals located at each vertex will operate using a simple set of update rules. These rules can incorporate random behavior, but the update decisions depend only on the color of an individual's neighbors. Consider the relationship between update rules for anti-coordination and coordination games. We will see that any update rule for an individual playing an anti-coordination game can be adapted to an update rule for playing a coordination game and vice versa. 
At its most basic, an anti-coordination rule aims to minimize the number of neighbors with the same color, and the goal of a coordination rule is to maximize the number of neighbors with the same color. Therefore, we can turn an anti-coordination update rule into a coordination update rule just by picking the opposite color every time.

\begin{figure}
    \centering
    \includegraphics[width = \textwidth]{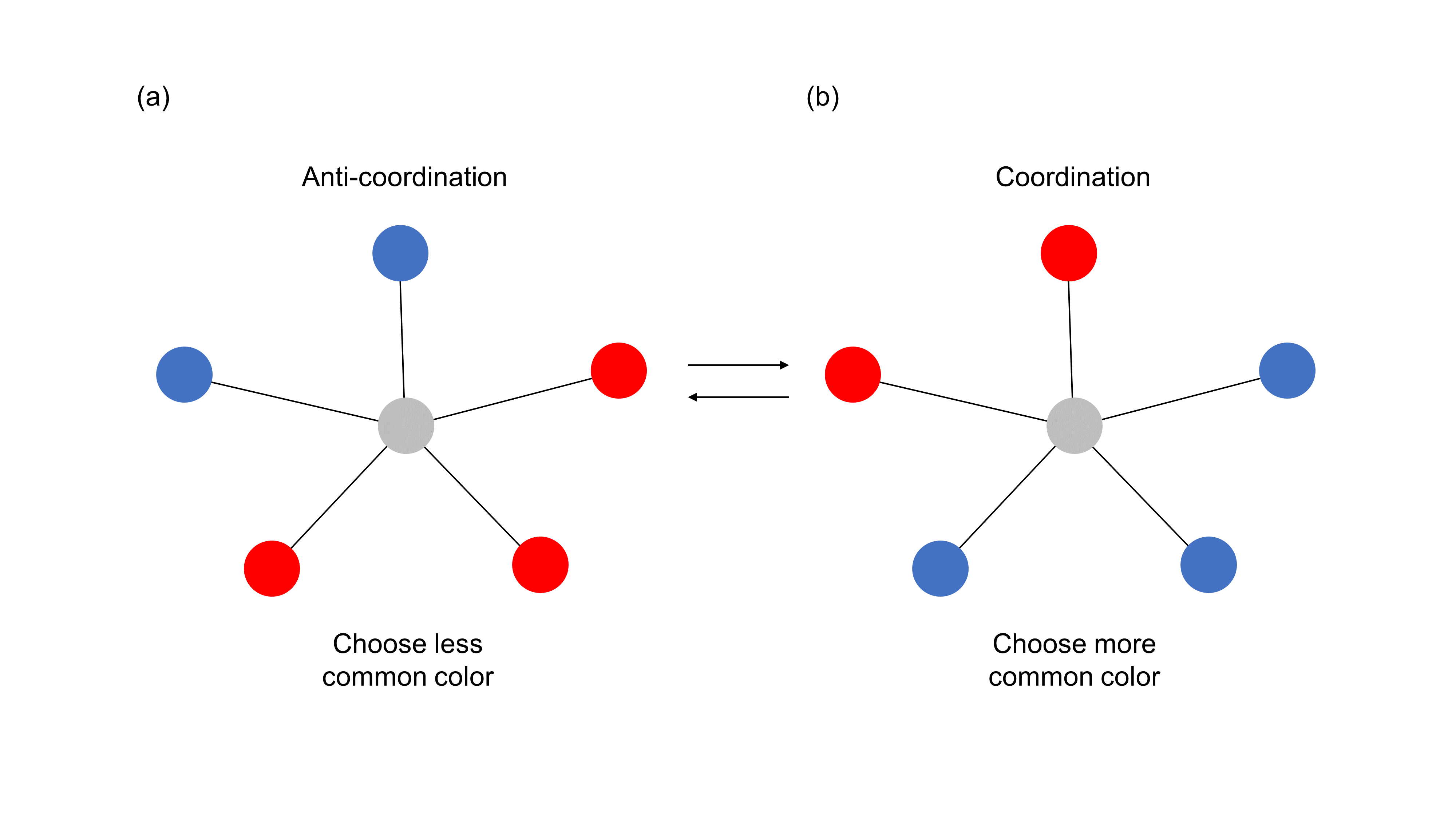}
    \caption{A simple case to demonstrate the bijection of update rules with two color choices. Making an anti-coordination decision in (a) will have the same outcome as making a coordination decision in (b), since all the colors of the neighbors have changed to the other color. If an individual would have chosen blue in (a) to match with as few neighbors as possible, that would correspond to choosing blue in (b), where the goal is to match with as many neighbors as possible.}
    \label{fig:simpleCase}
\end{figure}

Suppose we have an individual vertex with $a$ neighbors playing color A and $b$ neighbors playing color B (see Figure \ref{fig:simpleCase}a). When we define an anti-coordination rule where the central individual will select color A with probability $p(a,b)$ and color B with probability $1-p(a,b)$, we can make the corresponding coordination rule as follows: choose A with probability $1-p(a,b)$ and B with probability $p(a,b)$. 

Consider an update rule (anti-coordination or coordination) that has a function $p(a,b)$ that gives the probability of choosing color A. If we switch the colors of all neighbors, the probability of choosing $A$ is now $p(b,a)$ because now $b$ neighbors are playing $A$ and $a$ neighbors are playing B. There is a natural restriction to impose on the possible update rules. If we switch the color of every neighbor (see Figure \ref{fig:simpleCase}), the probabilities of the central vertex choosing color A, $p(a,b)$, and color B, $1-p(a,b)$, should switch as well. This restriction gives us the following complementary condition, by setting the probability of choosing A equal to the probability of choosing B after switching all the neighbors' colors:

\begin{equation}\label{eq:symcond}
    p(a,b) = 1 - p(b,a)
\end{equation}

For any anti-coordination update rule, a vertex with $a$ A neighbors and $b$ B neighbors will choose A with some probability $p(a,b)$. If we switch the colors of all the neighbors, the vertex will choose A with probability $p(b,a)=1-p(a,b)$, but this is equal to the probability of a coordination player choosing A. 
Therefore, an anti-coordination algorithm can be converted into its dual algorithm for a coordination game by temporarily switching the colors of all the neighbors, using the anti-coordination update rule, and switching the neighbors' colors back. 
As an example, an anti-coordination update rule on Figure \ref{fig:simpleCase}(a) will have the same behavior as a coordination update rule on Figure \ref{fig:simpleCase}(b).

The same process can be used to convert a coordination algorithm to an anti-coordination algorithm.

To put the above individual choice function $p(a,b)$ in context, it is worthwhile to introduce a few intuitive anti-coordination update rules used in prior work on network graph coloring problems  \cite{jones2021}. The first update rule, called randomness-first, involves making a random choice with probability $r$, and otherwise with probability $1-r$ makes a color choice that minimizes color conflicts. This update rule can be expressed as:

\begin{equation}
    p(a,b) = \begin{cases}
    1-\frac{1}{2} r &  a<b \\
    \frac{1}{2} & a=b \\
    \frac{1}{2} r & b<a
    \end{cases}
\end{equation}

Under the second update rule, called memory-0, individuals first attempt to choose any color that eliminates all color conflicts. If that is not possible, the individual chooses randomly with probability $r$ and otherwise with probability $1-r$ chooses the color minimizing conflicts with neighbors. In our terms, this algorithm is

\begin{equation}
    p(a,b) = \begin{cases}
    1 & a=0 \\
    1-\frac{1}{2}r &  0<a<b \\
    \frac{1}{2} & a=b \\
    \frac{1}{2}r & 0<b<a \\
    0 & b=0
    \end{cases}
\end{equation}

The third main update rule, called memory-1, is like the memory-0 rule except that the agent only makes a random choice if no neighbors have changed color in the last round of updates. Since this is not a memory-less update rule, it does not have a corresponding $p(a,b)$ function, and the following proof would need to be slightly modified, particularly by significantly enlarging the state space of the Markov chains to include the last $N$ colorings of the graph, to prove the equivalence for update rules with finite memory. While we do not go over all the details of proving that a finite-memory update rule also satisfies the isomorphism, we do show results of computer simulations to demonstrate that the duality of coordination and anti-coordination holds in Section 3.

In what follows, we demonstrate that an anti-coordination update rule is exactly as effective at finding a 2-coloring as the corresponding coordination update rule is at finding a uniform color for the whole bipartite network.

\subsection{Two Markov Chains}
For a connected, bipartite graph $G$ of size $N$, let $\textrm{col}(G)$ be the set of all possible labelings of the graph $G$. Note that here we refer to all ways of labeling the vertices of $G$ with either color A or color B, not just 2-colorings in which no neighbors share the same color.

The system will update as follows: the graph is initialized by randomly assigning each vertex a color. An update order is created that describes the order in which the labelled vertices will update their color. The update order is represented as a list of the numbers 1 through $N$, which is just a permutation of $N$ elements. The set of all permutations of $N$ elements, called the symmetric group on $N$ elements, is denoted $S_N$. The vertices continually update their colors in this order, one at a time, until the desired coloring (either a 2-coloring or uniform coloring) is found.

Now we can define our Markov chains. Let $\{X_i\}$ be a Markov chain using an anti-coordination update rule, and let $\{Y_i\}$ be the Markov chain using the associated coordination update rule, as described above. The state space $\Omega$ of both chains is the set of ordered triples $(G^*,\sigma,m)$ where $G^* \in \textrm{col}(G)$, $\sigma \in S_n$, and $m \in \{1,2,\dots, n\}$. Unsurprisingly, $G^*$ represents the colors of the vertices of the graph at some time $i$. $\sigma$ is the order in which the vertices update, and $m$ is the current position in the update step.

\begin{figure}
    \centering
    \includegraphics[width =\textwidth]{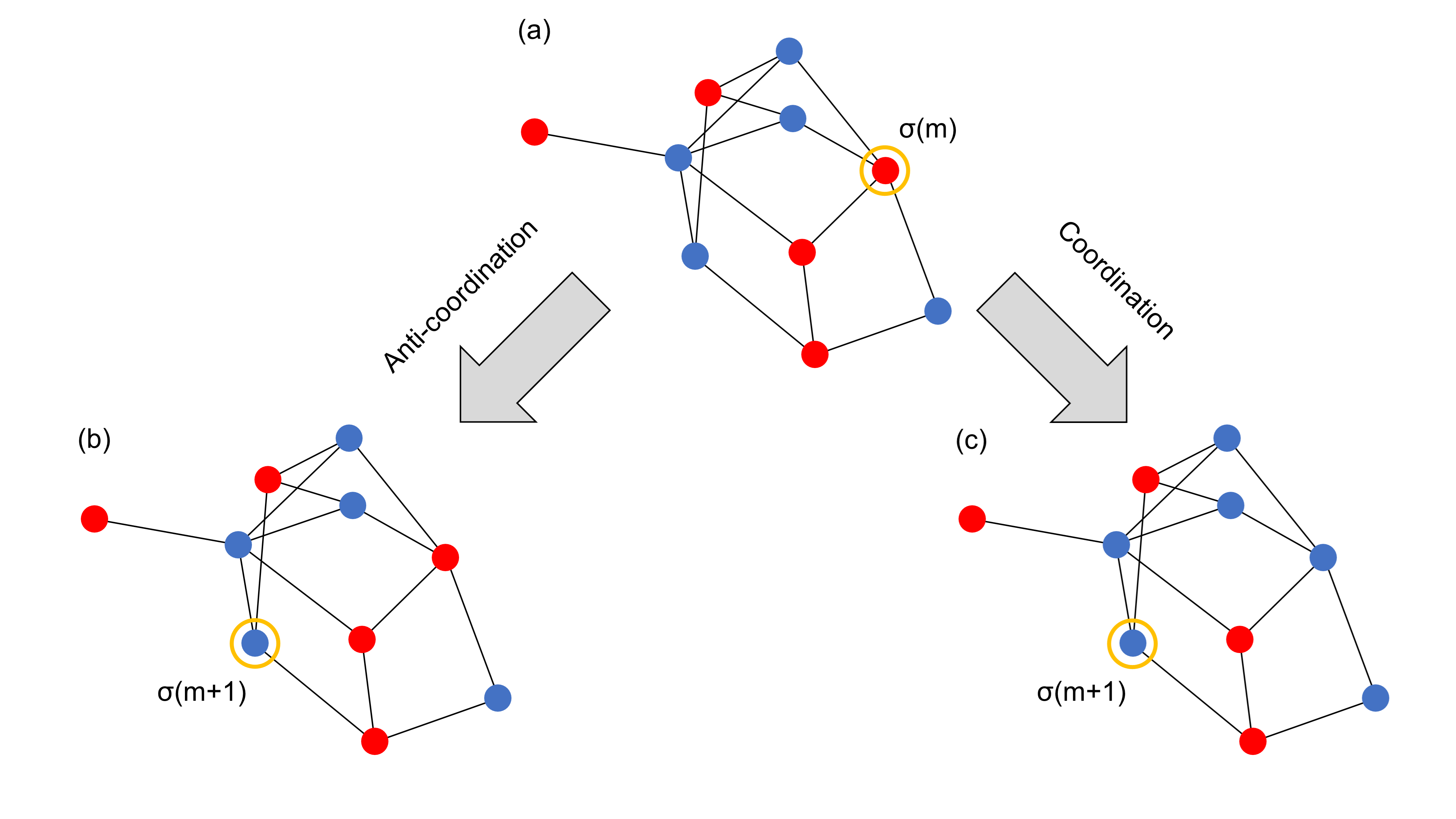}
    \caption{A demonstration of the possible transitions in both Markov chains. The next vertex to update is marked by a gold ring. Transitioning from (a) to (b) is minimizing matching with neighbors' colors, and is more likely to appear in an anti-coordination Markov chain, while transitioning from (a) to (c) is matching with as many neighbors as possible, and more likely in the coordination Markov chain.}
    \label{fig:transitions}
\end{figure}

The state space is quite large, but for each state, there are exactly two states to which the Markov chains can move with non-zero probability, shown in Figure \ref{fig:transitions}.

To begin, we initialize both Markov chains (anti-coordination and coordination) by sampling from the uniform distribution $\Pi$ over $\Omega$, so each starting coloring is equally likely.

Without loss of generality, let $X_j=Y_j=(G^*,\sigma,m)$. Here, $\sigma(m)$ is the vertex that is about to update. Let $G^*_A$ be the colored graph that is the same as $G^*$ except possibly $\sigma(m)$ which has color A, and $G^*_B$ the same but for color B. In each step of the Markov chains, $\sigma(m)$ selects one of two colors and the position in the update cycle increases by one, resetting to 1 if necessary. The update order $\sigma$ remains unchanged. Thus, if $\sigma(m)$ has $a$ color A neighbors and $b$ color B neighbors, 
\begin{equation}\label{eq:xa}
P(X_{j+1} = (G^*_A,\sigma,m\modn+1)) = p(a,b)\end{equation}
\begin{equation}\label{eq:xb}
P(X_{j+1} = (G^*_B,\sigma,m\modn+1)) = 1-p(a,b)=p(b,a)\end{equation}
\begin{equation}\label{eq:ya}
P(Y_{j+1} = (G^*_A,\sigma,m\modn+1)) = 1-p(a,b=p(b,a))\end{equation}
\begin{equation}\label{eq:yb}
P(Y_{j+1} = (G^*_B,\sigma,m\modn+1)) = p(a,b)\end{equation}

\subsection{A Markov Chain Isomorphism}
For bipartite graphs, we claim that these Markov chains $\{X_i\}$ and $\{Y_i\}$ are isomorphic. First, because $G$ is a connected, bipartite graph, the vertices can be divided into two groups. In a 2-coloring, all the vertices in the same group will be the same color, and all vertices in different groups will be different colors. Let $S$ be the set of vertices of one of these groups. Because we are working with 2-colorings of bipartite graphs, we can define a function $\phi:\textrm{col}(G) \to \textrm{col}(G)$ by switching the color of every vertex in $S$, and define $\psi_S:\Omega \to \Omega$ as the extension of $\phi$ in the natural way. We claim that this is a Markov chain isomorphism between $X_i$ and $Y_i$. This requires proving two conditions hold. First, $\psi_S$ must be bijective. Second, $\psi_S$ must commute with the transition matrices of $X_i$ and $Y_i$, i.e. the probability of $X_i$ moving from $x$ to $y$ is the same as $Y_i$ moving from $\psi_S(x)$ to $\psi_S(y)$. More formally, for all $x, y \in \Omega$, 
\begin{equation}\label{eq:isomorphism}
    P(X_{i+1} = y|X_i = x) = P(Y_{i+1} = \psi_S(y)|Y_i=\psi_S(x))
\end{equation}
If Equation \ref{eq:isomorphism} holds, the two Markov chains are equivalent in that after relabelling the states in $\Omega$ (according to $\psi_S$), the Markov chains are identical.

\subsection{Proof of isomorphism}

That $\psi_S$ is bijective is fairly obvious. For any colored graph $G^*$, because we are only working with 2-colorings on bipartite graphs, $\phi(G^*)$ is well-defined, and only $\phi(G^*)$ maps to $G^*$, so it is both one-to-one and onto, and therefore $\psi$ is as well.

Now we will prove Equation \ref{eq:isomorphism}. Since we are considering Markov chains moving from $x$ to $y$ (or $\psi_S(x)$ to $\psi_S(y)$), let $x = (G^*, \sigma, m)$. Let $a$ and $b$ be the number of color A and color B neighbors of $\sigma(m)$ in $G^*$, respectively. 

We begin with the conditional statement $X_i = x = (G^*, \sigma, m)$. Equations \ref{eq:xa} and \ref{eq:xb} give the only two possibles states of $X_{i+1}$ and their transition probabilities: 

\begin{equation}
P\bigg(X_{i+1} = (G^*_A,\sigma,m\modn+1)\bigg) = p(a,b)\end{equation}
\begin{equation}
P\bigg(X_{i+1} = (G^*_B,\sigma,m\modn+1)\bigg) = 1-p(a,b)\end{equation}

Once again, $G^*_A$ and $G^*_B$ are the same as $G^*$ except $\sigma(m)$ which has color A or B, respectively.

Now we consider $Y_{i+1}$ given that $Y_i = \psi(x) = \psi((G^*, \sigma, m)) = (\phi(G^*), \sigma, m)$. $\sigma(m)$ is the next vertex to update, and either it is in the subset $S$ or it is not. These two cases must be handled separately.

\subsection{Case 1: $\sigma(m) \in S$}

\begin{figure}
    \centering
    \includegraphics[width = \textwidth]{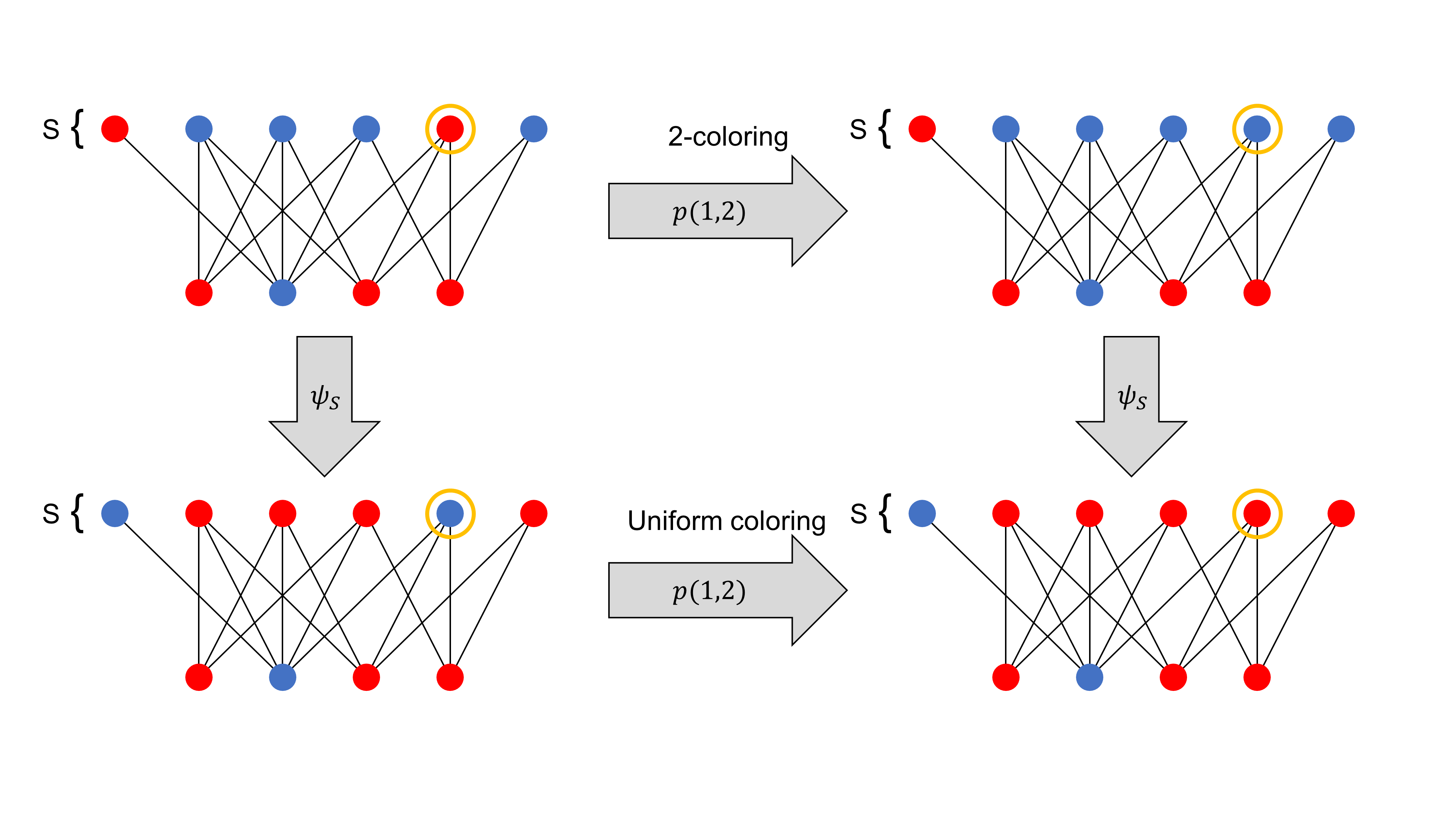}
    \caption{An example on a small bipartite graph showing that $\psi_S$ commutes with the transition matricies, when $\sigma(m) \in S$. Color A is blue and color B is red. The top row shows the transition in the anti-coordination Markov chain, and the bottom is the transition in the coordination Markov chain. In both chains, this particular transition occurs with probability $p(1,2)$.}
    \label{fig:sigmainS}
\end{figure}

If $\sigma(m) \in S$, none of $\sigma(m)$'s neighbors are in $S$, so $\sigma(m)$ still has $a$ color A neighbors and $b$ color B neighbors. Because we are now in the coordination Markov chain $\{Y_i\}$, $\sigma(m)$ chooses its color according to equations \ref{eq:ya} and \ref{eq:yb}.

With probability $p(a,b)$, $\sigma(m)$ chooses color B. Because $\sigma(m) \in S$, $\phi(G^*)$ becomes $\phi(G^*_A)$ when $\sigma(m)$ chooses B. Thus, $Y_{i+1}=(\phi(G^*_A),\sigma, m \modn+1) = \psi(G^*_A,\sigma,m\modn+1)$.

With probability $1-p(a,b)$, $\sigma(m)$ chooses color A, and $Y_{i+1} = (\phi(G^*_B),\sigma, m \modn+1) = \psi(G^*_B,\sigma,m\modn+1)$.

Thus, when $\sigma \in S$, Equation \ref{eq:isomorphism} holds.

\subsection{Case 2: $\sigma(m) \notin S$}

\begin{figure}
    \centering
    \includegraphics[width = \textwidth]{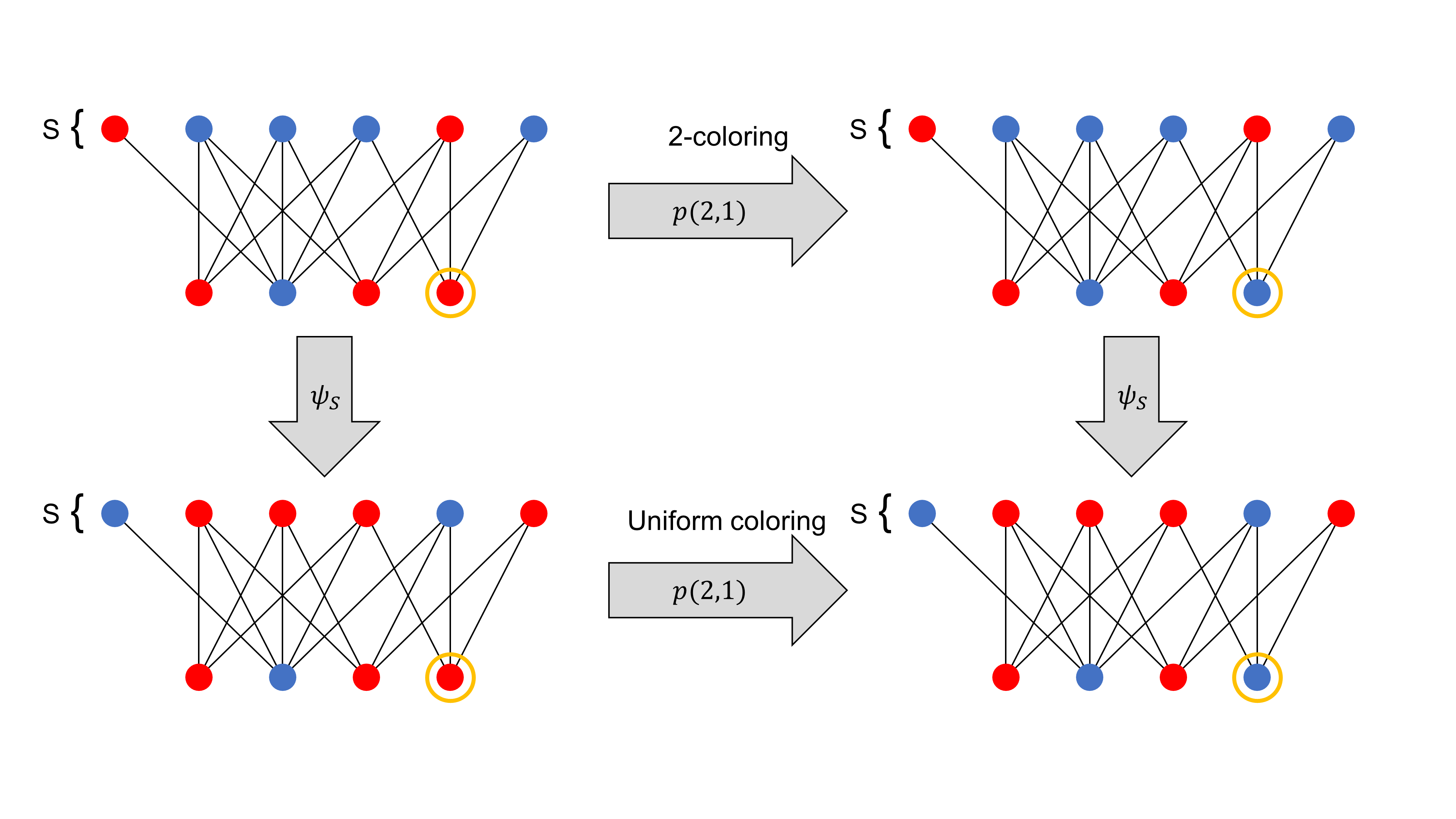}
    \caption{An example showing that $\psi_S$ commutes with the transition matricies when $\sigma(m) \notin S$. Color A is blue and color B is red. The top is the anti-coordination Markov chain, and the bottom is the coordination Markov chain. This time, the transition occurs with probability $p(2,1)$.}
    \label{fig:sigmanotinS}
\end{figure}

If $\sigma(m) \notin S$, then all of its neighbors are. So in $\phi(G^*)$, $\sigma(m)$ has $b$ color A neighbors and $a$ color B neighbors. 

With probability $1-p(b,a) = p(a,b)$, $\sigma(m)$ chooses color A, and $Y_{i+1} = (\phi(G^*_A), \sigma, m \modn+1)$.

With probability $p(b,a) = 1-p(a,b)$, $\sigma(m)$ chooses B, and $Y_{i+1} = (\phi(G^*_A), \sigma, m \modn+1)$.

So Equation \ref{eq:isomorphism} holds when $\sigma(m) \notin S$. Therefore, $\psi$ is a Markov chain isomorphism.

\subsection{Equivalence of the 2-coloring and uniform coloring problems}

Now we are prepared to state and defend the main claim of this work: when using local information, the anti-coordination and coordination problems are equivalent. Any result regarding the efficacy of an update rule $p(a,b)$ for an anti-coordination game can also be applied to a coordination game, and vice versa.

Because the initial distribution $\Pi$ is the uniform distribution and $\psi$ is bijective, $\psi(\Pi) = \Pi$ and both Markov chains begin from the same distribution. Furthermore, because $\psi$ switches the color of the set $S$, for any state $X_i$ in which a 2-coloring has been found, $\psi(X_i)=Y_i$ has a uniform coloring. 
Now use Equation \ref{eq:isomorphism} and get that for all $x \in \Omega$ and for all times $i$:
\begin{equation}\label{eq:result}
    P(X_i = x |X_0 \sim \Pi) = P(Y_i =\psi(x)| Y_0 \sim \psi(\Pi) = \Pi)
\end{equation}

Critically, this says that the probability of solving the anti-coordination problem in $i$ steps is the same as solving the coordination problem in $i$ steps, for all $i$. Additionally, the process is linked at each step, so the expected number of player color changes will be the same, for example. 

This result also holds for any update rules with finite memory. Any stochastic process whose transition probabilities only depend on a finite number of previous states can be reexpressed as a Markov chain by defining the new state space to be lists of elements from the previous state space, and this works here with any update rule that considers the last $n$ update steps.

\section{Simulation Results}

This result has been confirmed with a variety of simulation results. First, we take a broad approach: We create a large number of different networks, and populating each with individuals playing a particular anti-coordination update rule. Then we repeatedly attempt to find a 2-coloring of the network, collecting data on probability of finding a 2-coloring, the number of update cycles needed, and the number of players updated. Then, using the same network with individuals playing the associated coordination update rule, we repeatedly search for a uniform coloring, collecting data on the same metrics. After repeating this on all the networks, we have a large data collection that, if anti-coordination and coordination games are equivalent, should be two samples of the same probability distribution. 

And we see that this is the case using the two-sample Kolmogorov-Smirnov test on data collected from 1,000 different networks. For all three metrics (probability of solving the network, update cycles, and updated players), the K-S statistic is below 0.015 with a p-value greater than 0.999. This strongly suggests that the samples are drawn from the same distribution and the two problems are equivalent.

\begin{figure}
    \centering
    \includegraphics[width=\textwidth]{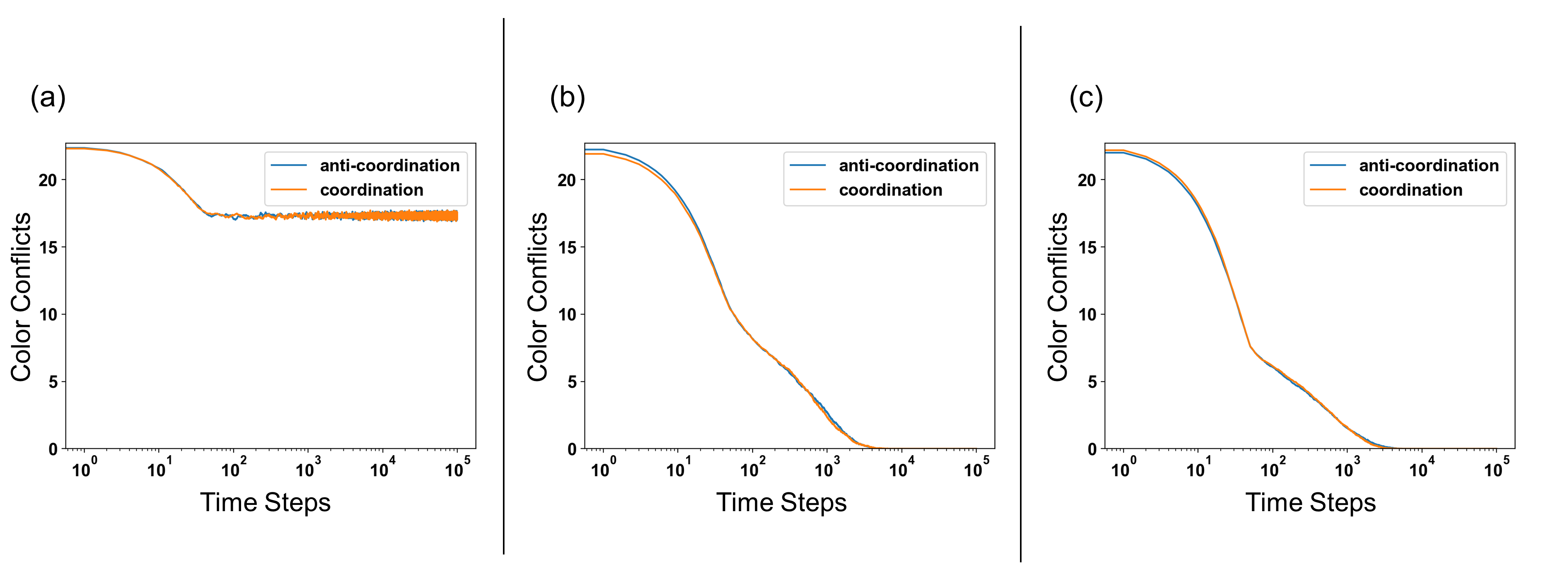}
    \caption{Plots showing the time evolution of the number of color conflicts using three reasonable update rules. (a) is randomness-first, (b) is memory0, and (c) is memory1. Crucially, the anti-coordination and coordination variants of the same update rule have the same behavior in all three plots. Curves are the average of 1000 simulations for each update rule. For randomness-first, the random behavior probability was 0.5. For Memory0 and Memory1, the random probability was 0.1.}
    \label{fig:conflictCounts}
\end{figure}

We can also consider a closer examination of the moment-to-moment behavior of each system by counting the number of color conflicts in the network at every time step, averaged over multiple runs. A color conflict is an edge who ends have the same color (in the case of an anti-coordination game) or different colors (in the case of a coordination game). Previous work \cite{jones2021} dealt mainly with three update rules: randomness-first, memory-0, and memory-1. In Figure \ref{fig:conflictCounts}, we see the result of many simulations on the same graph, with these three different update rules. The x axis is log scaled, to clearly show the behavior in the short and long term.

Although the proof given above doesn't strictly apply to the memory-1 update rule, it can be modified to work for any update rule that gives its agents finite memory by enlarging the state space to ordered tuples of network colorings. In Figure \ref{fig:conflictCounts}c, we see that finding uniform colorings and 2-colorings are equally difficult on random bipartite graphs.

These simulations confirm that the behavior when searching for a 2-coloring is the same as when searching for a uniform coloring, regardless of the specific update rule.

\section{Discussion \& Conclusion}

Studying the collective behavior of individuals in a large group has long been an important research area of statistical physics and relevant fields. The question of ``collective action,'' the tendency for individuals in a group to forgo short-term selfish behavior in favor of long-term group benefit, has been extensively discussed and examined. Of particular interest is classifying the environmental factors that foster cooperation within group, particularly in the case of a public goods game and the Prisoner's Dilemma \cite{hardin2015}. There are a plethora of studies that use networks to model a social structure on the group, and the exact topology of networks can have a profound impact on the cooperation inside a group \cite{pena2012, allen2019, szabo2010, perc2017}. Additionally, empirical research uses human trials to examine how humans behave rationally (or irrationally) when actually playing public goods games with others \cite{ostrom2000}.

Our results add to the study of collective action by approximating public goods games in that individuals sometimes need to make selfless actions (choosing colors that increase color conflicts) with the long-term goal of increasing success for the entire group (finding a 2-coloring or uniform coloring)~\cite{jones2021}. Our present work shows that these two fundamentally different games behave in the same way on random bipartite networks.

Our finding is counter intuitive, but it is important to remember that it applies in a relatively narrow range of scenarios. Anti-coordination and coordination are equivalent problems only in a bipartite populations with an initial coloring sampled uniformly from all possible colorings. There are also no parallels for $n$-colorings for $n>2$. 

Our finding is counter intuitive, but it is important to remember that it applies in a relatively narrow range of scenarios. A bipartite structure is unlikely in most social networks, which means anti-coordination and coordination are equivalent problems only in the small selection of populations that happen to be bipartite with an initial coloring sampled uniformly from all possible colorings. There are also no parallels for $n$-colorings for $n>2$. 

\section*{Acknowledgements}
F.F. is grateful for the generous financial support by the NIH COBRE Program (grant no. 1P20GM130454), the Bill \& Melinda Gates Foundation (award no. OPP1217336) and the Neukom CompX Faculty Grant.

\section{References}

\begin{thebibliography}{10}


\bibitem{kearns2006}
Michael Kearns and Siddharth Suri and Nick Montfort.
\newblock An Experimental Study of the Coloring Problem on Human Subject Networks.
\newblock{\em Science} 313(5788):824-827, 2006.

\bibitem{mccubbins2009}
Mathew ~D. McCubbins and Ramamohan Paturi and Nicholas Weller.
\newblock Connected Coordination: Network Structure and Group Coordination
\newblock{\em American Politics Research} 37(5):899-920, 2009.

\bibitem{jackson2014}
Matthew ~O. Jackson and Yiqing Xing.
\newblock Culture-dependent strategies in coordination games.
\newblock{\em Proceedings of the National Academy of Sciences} 111 (Supplement 3):10889--10896, 2014.

\bibitem{belloc2019}
Marianna Belloc and Ennio Bilancini and Leonardo Boncinelli and Simone D’Alessandro.
\newblock Intuition and Deliberation in the Stag Hunt Game.
\newblock{\em Scientific Reports}  9:14833, 2019.

\bibitem{mccain2014}
Roger ~A. McCain and Richard Hamilton.
\newblock Coordination games, anti-coordination games, and imitative learning.
\newblock{\em Behavioral and Brain Sciences} 37(1):90--91, 2014.

\bibitem{bramoulle2007}
Yann Bramoulle.
\newblock Anti-coordination and social interactions.
\newblock{\em Games and Economic Behavior} 58(1):30--49, 2007

\bibitem{neugebauera2008}
Tibor Neugebauera and Anders Poulsen and Arther Schramc.
\newblock Fairness and reciprocity in the Hawk-Dove Game.
\newblock{\em Journal of Economic Behavior \& Organization} 66(2):243--250, 2008.

\bibitem{sabar2012}
Nasser ~R. Sabar, Masri Ayob, Rong Qu, Graham Kendall.
\newblock A graph coloring constructive hyper-heuristic for examination timetabling problems.
\newblock{\em Applied Intelligence}, 37(1):1--11, 2012.

\bibitem{park1996}
Taehoon Park and Chae ~Y. Lee.
\newblock Application of the graph coloring algorithm to the frequency assignment problem.
\newblock{\em Journal of the Operations Research society of Japan} 39(2): 258--265, 1996.

\bibitem{shirado2017}
H. Shirado and N. Christakis.
\newblock Locally noisy autonomous agents improve global human coordination in network experiments. 
\newblock{\em Nature} 545:370--374, 2017.

\bibitem{jones2021}
Matthew Jones and Scott Pauls and Feng Fu.
\newblock Random Choices Facilitate Solutions to Collective Network Coloring Problems by Artificial Agents.
\newblock{\em iScience} 24(4), 2021.

\bibitem{broere2017}
Joris Broere and Vincent Buskens and Jeroen Weesie and Henk Stoof.
\newblock Network effects on coordination in asymmetric games.
\newblock{\em Scientific Reports} 7: 17016, 2017.

\bibitem{broere2019}
Joris Broere and Vincent Buskens and Henk Stoof and Angel Sanchez.
\newblock An experimental study of network effects on coordination in asymmetric games.
\newblock{\em Scientific Reports} 9: 6842, 2019.

\bibitem{mccubbins2020}
Mathew McCubbins and Nicholas Weller.
\newblock Coordination, Communication, and Information: How Network Structure and Knowledge Affect Group Behavior.
\newblock{\em Journal of Experimental Political Science} 7:1--12, 2020.

\bibitem{choi2011}
Syngjoo Choi and Douglas Gale and Shachar Kariv and Thomas Palfrey.
\newblock Network architecture, salience and coordination.
\newblock{\em Games and Economic Behavior} 73(1):76--90, 2011.

\bibitem{shirado2020}
Hirokazu Shirado and Nicholas Christakis.
\newblock Network Engineering Using Autonomous Agents Increases Cooperation in Human Groups.
\newblock{\em iSicence} 23(9):101438, 2020.

\bibitem{rand2011}
David Rand and Samuel Arbesman and Nicholas Christakis.
\newblock Dynamic social networks promote cooperation in experiments with humans.
\newblock{\em Proceedings of the National Academy of Sciences} 108(48):19193--19198, 2011.

\bibitem{szabo2007}
Gy\"{o}rgy Szab\'{o} and G\'{a}bor F\'{a}th.
\newblock Evolutionary games on graphs.
\newblock{\em Physics Reports} 446(4-6):97--216, 2007.

\bibitem{gagniuc2017}
Paul Gagniuc.
\newblock Markov Chains: From Theory to Implementation and Experimentation. 
\newblock{\em John Wiley \& Sons} USA, NJ 1--165, 2017.

\bibitem{durrett2018}
Richard Durrett.
\newblock Essentials of Stochastic Processes.
\newblock{\em Springer International Publishing}, 2018.

\bibitem{pena2012}
Jorge Pena and Yannick Rochat.
\newblock Bipartite graphs as models of population structures in evolutionary multiplayer games.
\newblock{\em Public Library of Science San Francisco} USA, 2012.

\bibitem{ostrom2000}
Elinor Ostrom.
\newblock Collective action and the evolution of social norms.
\newblock{\em Journal of economic perspectives} 14(3):137--158, 2000.

\bibitem{hardin2015}
Russell Hardin.
\newblock Collective action.
\newblock{\em RFF Press} 2015.

\bibitem{allen2019}
Benjamin Allen and Gabor Lippner and Martin Nowak.
\newblock Evolutionary games on isothermal graphs.
\newblock{\em Nature communications} 10(1):1--9, 2019.

\bibitem{szabo2010}
Gyoergy Szabo and Attila Szolnoki and Melinda Varga and Livia Hanusovszky.
\newblock Ordering in spatial evolutionary games for pairwise collective strategy updates.
\newblock{\em Physical Review E} 82(2):262110, 2010.

\bibitem{perc2017}
Matja{\v{z}} and Jillian J Jordan and David G Rand and Zhen Wang and Stefano Boccaletti and Attila Szolnoki.
\newblock Statistical physics of human cooperation.
\newblock{\em Physics Reports} 687:1--51, 2017.



\end{thebibliography}

\end{document}